\pdfoutput=1
\documentclass[final,5p,times,twocolumn]{elsarticle}

\usepackage{algorithm}
\usepackage{algorithmicx}
\usepackage{algpseudocode}
\usepackage{amsmath}
\usepackage{graphicx}
\usepackage{epstopdf}
\usepackage{amsthm}
\usepackage{graphicx}
\usepackage{subfigure}
\usepackage{float}
\usepackage{color}
\usepackage{amsmath }
\usepackage{amsfonts}
\usepackage{graphicx}
\usepackage{arydshln}
\usepackage{verbatim}
\usepackage{subfigure}
\usepackage{enumerate}
\usepackage{rotating}
\usepackage{threeparttable}
\usepackage{caption}            
\biboptions{sort&compress}

\theoremstyle{remark}

\begin{document}

\begin{frontmatter}

\title{Vapor Compression Cycle Control for Automotive Air Conditioning Systems \\with a Linear Parameter Varying Approach}

\author{Xu Zhang}\ead{xuzhanghit98@gmail.com}

\address{Department of Control Science and Engineering, Harbin Institute of Technology, Harbin, 150080, China.}

\begin{abstract}
This paper investigates an output tracking problem for the vapor compression cycle in automotive Air Conditioning (A/C) systems using Linear Parameter Varying (LPV) techniques. Stemming from a recently developed first-principle A/C model, Jacobian linearization is first exploited to develop an LPV-based model that is nonlinearly dependent on time-varying system parameters such as evaporator pressure and superheat temperature. To facilitate the control implementation, a Tensor Product (TP) model transformation is applied to transform the LPV-based model to a TP-type convex polytopic model. LPV controllers are then designed to guarantee system stability, robustness and H-infinity performance. Simulations are presented to demonstrate the efficacy of the developed framework.
\end{abstract}

\begin{keyword}
Air Conditioning System \sep Optimization \sep Linear Parameter Varying Control \sep Moving Boundary Method \sep Tensor-Product Transformation
\end{keyword}

\end{frontmatter}

\section{Introduction}
During the past decade, significant research emphasis has been given to the development of advanced technologies for the automotive industry. However, the control of the automotive Air Conditioning (A/C) is still considered as an open problem and needs to be investigated. Reduction on energy consumption of AC systems can be realized through model-based optimization and control designs. In general, control design for A/C and refrigeration systems involves a formulation of output tracking on evaporator pressure and superheat temperature \cite{04_zhang2013, 04_rasmussen2010}. To enable a model-based control, dynamic A/C models are essential to accurately predict  pressure and enthalpy changes in the evaporator and condenser. Recently, the Moving Boundary Method (MBM) has been a widely applied technique to model the pressure dynamics \cite{01_Asada1998, 02_He1997, 02_Li2010, 02_McKinley2006}.

 However, MBM-based models are typically in forms of high-order Nonlinear Differential and Algebraic Equations (NDAEs), which impedes system control design. To resolve this complication,  lower-order linear models, obtained through system identification or model order reduction \cite{04_zhang20132, 02_Jensen2003,J6}, are frequently used. For instance, in \cite{04_shah2004}, a discrete-time state variable model for indirect adaptive control is recursively identified using a multi-input multi-output (MIMO) parameter estimation algorithm. A linear quadratic regulator (LQR) is then implemented for reference tracking and disturbance rejection. In \cite{04_leducq2006}, a simplified low-order model is used in the predictive part of an MPC algorithm by minimizing the weighted sum of three quadratic partial criteria for a chiller, while the superheat value is controlled separately by a PID controller.  Similarly, a low-order nonlinear evaporator model is developed for backstepping design of a nonlinear adaptive controller \cite{04_rasmussen2011}.

These simplified modeling and control frameworks, however, are only capable of achieving local output tracking, their extension to global tracking performance is constrained. As far as the authors are concerned, the main reasons behind limited literature on global output tracking design for A/C systems are twofold: firstly, it lacks a systematic treatment of the control-oriented A/C model to facilitate control algorithm design; secondly, the control theory adopted is typically \emph{ad hoc}, lacking theoretical foundation of stability and robustness.

In this paper, we exploit a linear parameter varying (LPV) control method that provides a framework not only providing a systematic way to represent general nonlinear model in a special form, but also guaranteeing system stability, robustness, and performance of the closed-loop system. Specifically, a Jacobian-based LPV model is derived from a nonlinear dynamic model of a vapor compression cycle. The TP model transformation is applied to transform the Jacobian-based LPV model into a TP-type convex polytopic model form. Since $H_{\infty}$ techniques have been demonstrated to be promising for both theoretical and industrial problems \cite{Du2005981,J9,C4,Yamashita19941717,C5}, we exploit the $H_{\infty}$ gain-scheduling control developed in \cite{50_wu1996, 50_apkarian1998} to synthesize an LPV controller that achieves the desired closed-loop properties based on the TP polytopic model.

\begin{figure*}[!htb]
\centering
\includegraphics[width=0.75\textwidth]{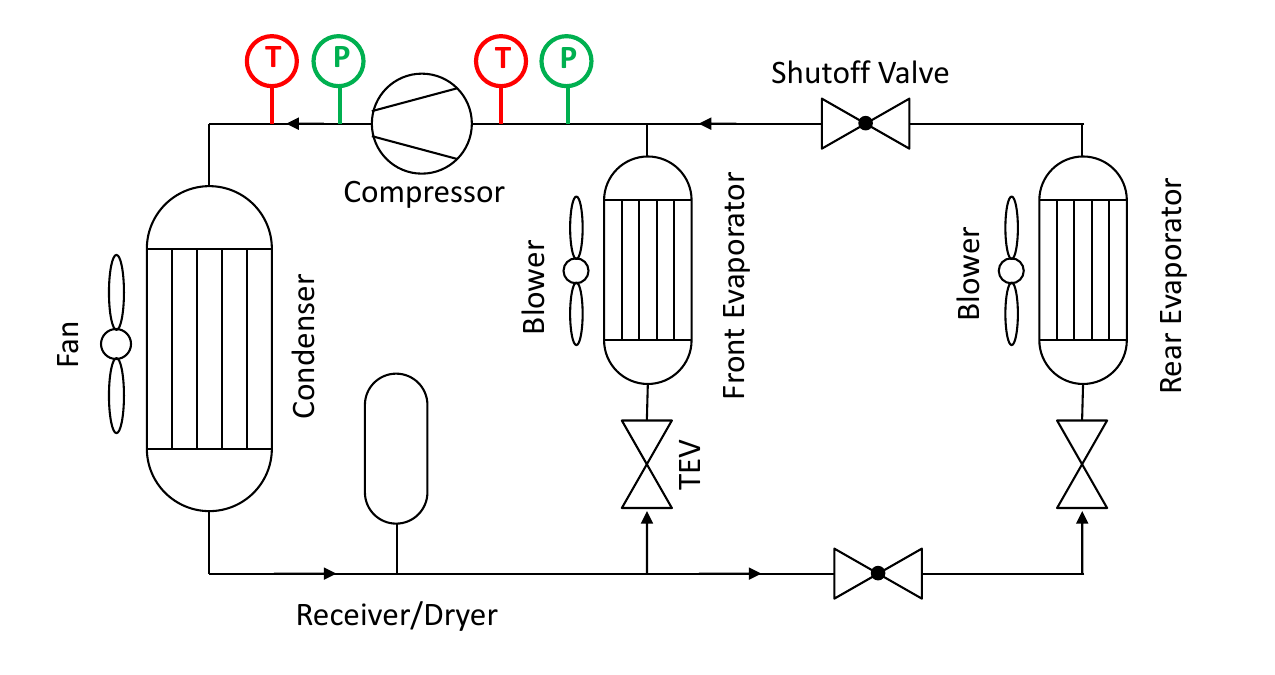}
\caption{Layout of the A/C System.}
\label{fig:AC_scheme}
\end{figure*}

The rest of this paper is organized as follows. Section 2 presents the control-oriented nonlinear A/C model. 
Section 3 illustrates the LPV model.
The proposed LPV-based model using using Jacobian linearization and tensor product decomposition is given in Section 4.
Self-scheduled $H_{\infty}$ controller is synthesized in Section 5, and the global output tracking performance is demonstrated.
Open issues are discussed in Section 6.
Finally, a conclusion is drawn in Section 7.

\section{Control-oriented A/C Model}

As illustrated in Figure \ref{fig:AC_scheme}, a basic automotive A/C system consists of four main components: evaporator, compressor, condenser and expansion valve. The enthalpy, mass flow rate, and pressure are exchanged between the four components. Basically, the two heat exchangers (evaporators) set the pressures of the system, while the compressor and expansion valve determine the mass flow rates at the inlet and outlet of the evaporator and condenser.

As the main actuators, the compressor and the expansion valve regulate the pressure and enthalpy in the A/C system. The heat and dynamics in the compressor and the expansion valve are generally modeled as static components. However, at the inlet and outlet of the heat exchanger, the refrigerant boundary conditions, i.e., mass flow rate and enthalpy, are provided by the compressor and expansion valve. As a result, models of the compressor and the expansion valve should be considered to describe the mass flow rate and enthalpy change across control devices.

In the compressor, mass flow rate, $\dot m_c$, and outlet enthalpy, $h_2$, are defined respectively as:
\begin{equation}
\label{E:CMP_E}
\begin{aligned}
   \dot m_c &= \eta_v V_d \rho_1 \omega_c ,\\
    h_2 &= \frac {h_{2s} - h_1}{ \eta_s} + h_1,
   \end{aligned}
\end{equation}
where $V_d$ is the compressor displacement; $\rho_1$ and $h_1$ are the refrigerant density and enthalpy at the compressor inlet, respectively; $\omega_c$ is the compressor speed and $h_{2s}-h_1$ is the isentropic enthalpy difference. The first control input is the compressor rotation speed $N_c$ with the unit of $rpm$.

The mass flow rate through the expansion valve is modeled by the orifice flow equation, approximated by assuming constant fluid density:
\begin{equation}
\label{E:TEVM}
  \dot m_v = C_{d,v} A_v \sqrt{2\rho_3\left(p_{3} - p_{4}\right)},
\end{equation}
where $A_v$ is the valve curtain area and $C_v$ is the discharge coefficient. The outlet enthalpy is typically obtained by assuming an ideal throttling process, hence $h_4 = h_3$. The second control input is the valve position $\alpha$ in percentage, determining the effective flow area of the valve.

The mass and energy balance equations for the two-phase region and superheated region of the evaporator are given in Equations \ref{E:evap_31} and \ref{E:evap_41}, respectively. In these differential equations, the left hands represent the variation of independent states of the refrigerant, and the right hands characterize the exchange of mass and energy at the inlet and outlet of individual phase region, as well as the heat transfer along the wall of corresponding regions. The terms multiplying the state variations depend on the refrigerant inherent thermodynamic properties, hence are state-dependent. The mass and energy balances for the sub-cooled, two-phase and superheated region of the condenser are not shown here for brevity. Readers can refer to \cite{02_He1997,02_Li2010,50_zhang2014JDSMC} for detailed derivations.

\begin{figure*}
    \centering
      \subfigure[Vehicle Speed Profile]
      {\includegraphics[width=0.40\textwidth]{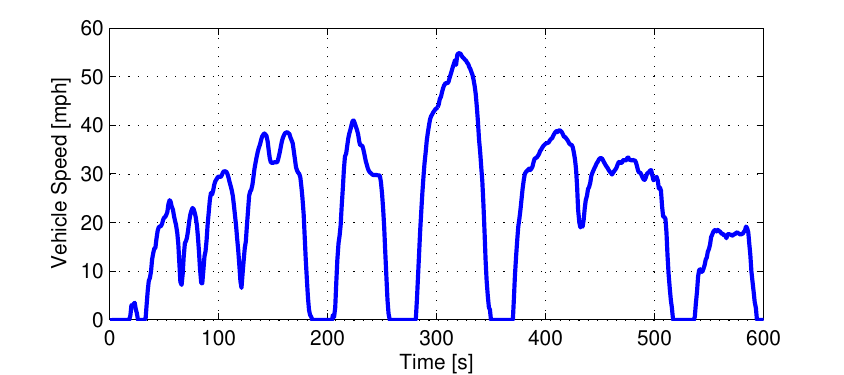}
      \label{fig:Veh_SC03}}
      \subfigure[Condenser Pressure]
      {\includegraphics[width=0.40\textwidth]{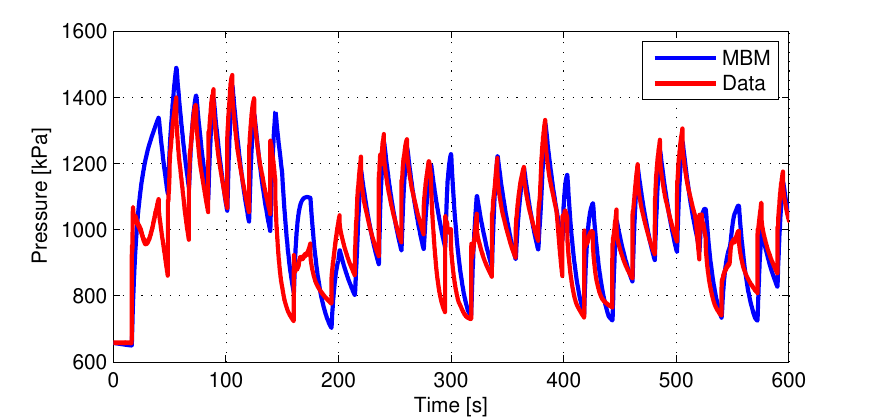}
      \label{fig: Pc_SC03}}
      \subfigure[Evaporator Pressure]
      {\includegraphics[width=0.40\textwidth]{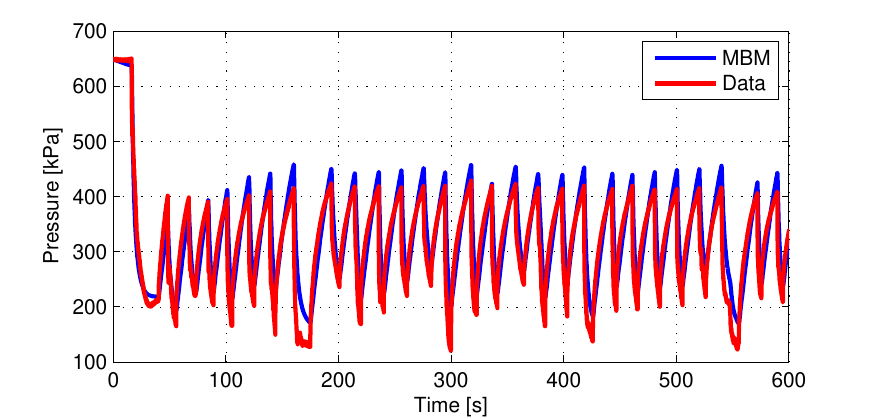}
      \label{fig: Pe_SC03}}
      \subfigure[Evaporator Exit Temperature]
      {\includegraphics[width=0.40\textwidth]{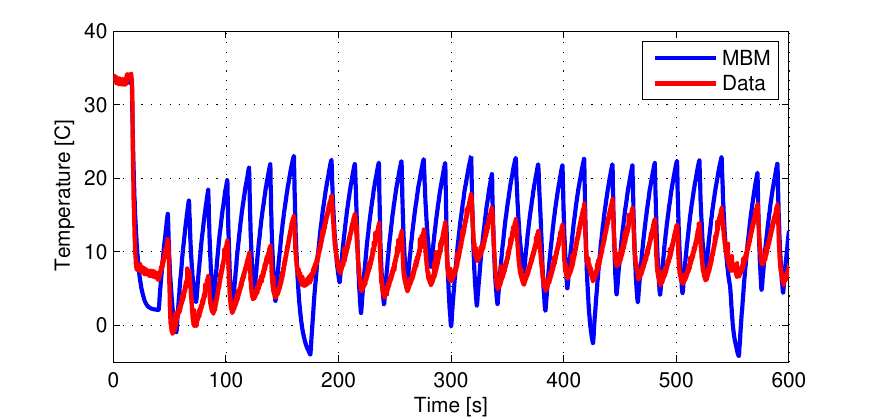}
      \label{fig:SC03_SH}}
      \caption{Verification of MBM and EBM for the SC03 driving cycle.}
      \label{fig: Val_SC03}
\end{figure*}

\begin{equation}
\label{E:evap_31}
\begin{aligned}
  &\left(\frac{\rho_{e,TP}-\rho_g}{\rho_{e,TP}}\right)\frac{d\zeta_1}{dt} + \frac{1}{\rho_{e,TP}}\frac{\partial \rho_{e,TP}}{\partial p_e}\frac{dp_e}{dt}\cdot\zeta_1\\
  &\qquad + \frac{1}{\rho_{e,TP}}\frac{\partial \rho_{e,TP}}{\partial \bar\gamma_e}\frac{d\bar\gamma_e}{dt}\cdot\zeta_1 \\
  &= \frac{\dot m_v}{\rho_{e,TP}V_e} - \frac{\dot m_{12}}{\rho_{e,TP}V_e}\cdot\frac{\rho_g \left(h_{e,TP}-h_g\right)}{\rho_{e,TP}}\frac{d\zeta_1}{dt}\\
   &\qquad+ \left(\frac{\partial h_{e,TP}}{\partial p_e}-\frac{1}{\rho_{e,TP}}\right)\frac{dp_e}{dt}\cdot\zeta_1+ \frac{\partial h_{e,TP}}{\partial \bar\gamma_e}\frac{d\bar\gamma_e}{dt}\cdot\zeta_1  \\
  & = \frac{\dot m_v}{\rho_{e,TP}V_e} \left(h_4-h_{e,TP}\right) - \frac{\dot m_{12}}{\rho_{e,TP}V_e} \left(h_g-h_{e,TP}\right) +\frac{\dot{Q}_{TP}}{\rho_{e,TP}V_e}.
\end{aligned}
\end{equation}
\begin{equation}
\label{E:evap_41}
\begin{aligned}
  &-\left(\frac{\rho_{e,SH}-\rho_g}{\rho_{e,SH}}\right)\frac{d\zeta_1}{dt} + \frac{1}{\rho_{e,SH}}\frac{\partial \rho_{e,SH}}{\partial p_e}\frac{dp_e}{dt}\cdot\left(1-\zeta_1\right)\\
   &\qquad+ \frac{1}{\rho_{e,SH}}\frac{\partial \rho_{e,SH}}{\partial h_{e,SH}}\frac{dh_{e,SH}}{dt}\cdot\left(1-\zeta_1\right) \\
  &= \frac{\dot m_{12}}{\rho_{e,SH}V_e} - \frac{\dot m_{c}}{\rho_{e,SH}V_e}-\frac{\rho_g \left(h_{g}-h_{e,SH}\right)}{\rho_{e,SH}}\frac{d\zeta_1}{dt}\\
   &\qquad+ \frac{1}{\rho_{e,TP}}\frac{dp_e}{dt}\cdot\left(1-\zeta_1\right) - \frac{dh_{e,SH}}{dt}\cdot\left(1-\zeta_1\right)  \\
  & = \frac{\dot m_{12}}{\rho_{e,SH}V_e} \left(h_g-h_{e,SH}\right) - \frac{\dot m_{c}}{\rho_{e,SH}V_e} \left(h_1-h_{e,SH}\right) +\frac{\dot{Q}_{SH}}{\rho_{e,SH}V_e}.p
\end{aligned}
\end{equation}

%
%

The inputs are the compressor rotation speed and expansion valve opening percentage, i.e., $u=\begin{bmatrix} N_{c} & \alpha \end{bmatrix} ^T$. The boundary conditions are the variables describing the air side of the heat exchangers, and are treated as unknown disturbances, i.e., $ v =  \begin{bmatrix}\dot{m}_{ea}  & T_{ea,in}  \end{bmatrix} ^T $. The state vector describing the evaporator status includes 6 states as: $ x_e= \begin{bmatrix} \zeta_{e1} & p_e & h_{e2} & T_{e1w} & T_{e2w}  \end{bmatrix} ^T $.  Finally, the outputs are the pressures and superheat temperature,i.e., $y =\begin{bmatrix} p_e & p_c & T_{r,eo} \end{bmatrix}^T$. The $Z$ matrix and $f$ vector are complex expressions of refrigerant properties, heat transfer coefficients and geometric parameters \cite{02_He1997,02_Li2010,50_zhang2014JDSMC}.

Figure \ref{fig: Val_SC03} illustrates the comparison of the model outputs with the corresponding experimental data. During the SC03 test, the compressor speed (related to the engine speed) changes considerably, causing significant variations in the refrigerant flow rate that affect the pressure dynamics in the heat exchangers. This is particularly evident by observing the fluctuations of the condenser pressure, as shown in Figure \ref{fig: Pc_SC03}. The model captures the dynamics induced by the compressor speed and the on-off cycling of the clutch.

\section{Linear Parameter Varying (LPV) Control Design}
Among a variety of LPV synthesis algorithms, it is desirable to balance control performance and computation complexity from the very beginning of choosing appropriate LPV models. Basically, two types of models have been introduced for control synthesis and analysis purpose, i.e. grid LPV model and affine LPV model (or polytopic LPV model). The direct application of algebraic manipulation on the control-oriented A/C model yields a grid LPV model that is nonlinearly dependent on the time-varying parameters. LPV control theory states that linear matrix inequality (LMI) constraints have to be evaluated at all grid points, yielding an infinite number of LMIs to be solved. In contrast, the system matrices of the affine LPV model are known functions and depend affinely on the time-varying parameters varying in a polytope of vertices. Hence LMI constraints have to be evaluated only at all vertices points, yielding a finite number of LMIs to be solved \cite{50_chumalee2009,50_rangajeeva2011}. However, the derivation of an affine LPV model from the control-oriented A/C model is not straightforward.  To overcome this difficulty, an alternative is to use a tensor-product (TP) model transformation. As explained in \cite{50_petres2006,50_petres2007}, it uses a higher order singular value decomposition (HOSVD) in order to decompose a given N-dimensional tensor into a full orthonormal system in a special order of higher order singular values. After the decomposition process, a TP-type convex polytopic model suitable for control algorithm development \cite{50_wu1996, 50_apkarian1998} is obtained from a grid LPV model, where the parameter-dependent weighting functions of the LTI vertices components of the polytopic model are one-dimensional functions of the elements of the parameter vector of the original grid LPV model.

A grid LPV model is obtained after linearization of nonlinear system, which is given as follows:
\begin{equation}\label{E:statespace}
\begin{aligned}
    \dot{x}(t) & = A(p(t))x(t) +  B(p(t))u(t)        \\
    y(t)       & = C(p(t))x(t) +  D(p(t))u(t)
\end{aligned}
\end{equation}
where $t \in R$ denotes time,
$x\in R^n$ represents the state vector,
$y \in R^q$ is the output vector,
$u \in R^m$ is used to denote the control input vector,
and $p(t)$ is a time-varying parameter vector whose dimension is $N$.
Note that  vector $p(t)$ is unknown a \emph{priori} but can be measured online.
Moreover, we assume that $p(t)$
is in some set bounded, which has a known upper bound and a known lower bound, i.e. $p_1(t) \in [a_1,b_1], p_2(t) \in [a_2,b_2], \dots, p_N(t) \in [a_N,b_N]$.
Therefore, $S(p(t))$, which is the system matrix,
is a function of all time-varying parameters;  we write it by
\begin{equation}\label{E:grid}
    S(p(t)) = \left(
                \begin{array}{cc}
                  A(p(t)) & B(p(t)) \\
                  C(p(t)) & D(p(t)) \\
                \end{array}
              \right)
              \in R^{(n+q) \times (n+m)}
\end{equation}
The exogenous parameter $\rho(t)$ is unknown a priori but can be measured or estimated online. If the scheduling parameter is endogenous to the state dynamics, e.g., $\rho(t)$ is a state itself as it will be case for the A/C model, the system shall be called quasi-LPV.

Using the state-space form in Equation \ref{E:statespace}, the nonlinear A/C plant model has been written into grid LPV form in Equation \ref{E:grid}. As suggested in \cite{50_petres2006,50_petres2007}, the design load using a grid LPV model is more tedious than the one using an affine LPV model.
When an affine LPV model is written in a state-space form in Equation \ref{E:statespace},
we know that  $S(p(t))$, which is the system matrix, depends affinely on the time-varying parameters.
Also, we define the matrix polytope as the convex hull of a finite number of matrices with the same dimensions.
Therefore, we have

\begin{equation}\label{E:affine}
\begin{aligned}
    S_r(p(t))  & \in
    \left \{ \left (
      \begin{array}{cc}
        A(p_i) & B(p_i) \\
        C(p_i) & D(p_i) \\
      \end{array}
    \right )
    : i = 1 , \dots , r = 2^N \right \} ,\\
    S_r(p(t)) & = \sum_{i=1}^r \alpha_i(p(t)) S_i\\
             & = \sum_{i=1}^r \alpha_i(p(t))
              \left (
                    \begin{array}{cc}
                    A(p_i) & B(p_i) \\
                    C(p_i) & D(p_i) \\
              \end{array}
              \right ),
\end{aligned}
\end{equation}
where $p_i$ denotes the frozen time-varying parameters at each vertex, $r$ is the total number of vertices, $\alpha_i(p(t))$ are the weighting functions that are linearly dependent on the time-varying parameters such that $\alpha_i(p(t)) \in [0,1]: \sum_{i=1}^r \alpha_i(p(t))=1$, and $S_i$ are the LTI system matrices at each vertex.

\begin{figure}
\centering
\includegraphics[width=0.5\textwidth]{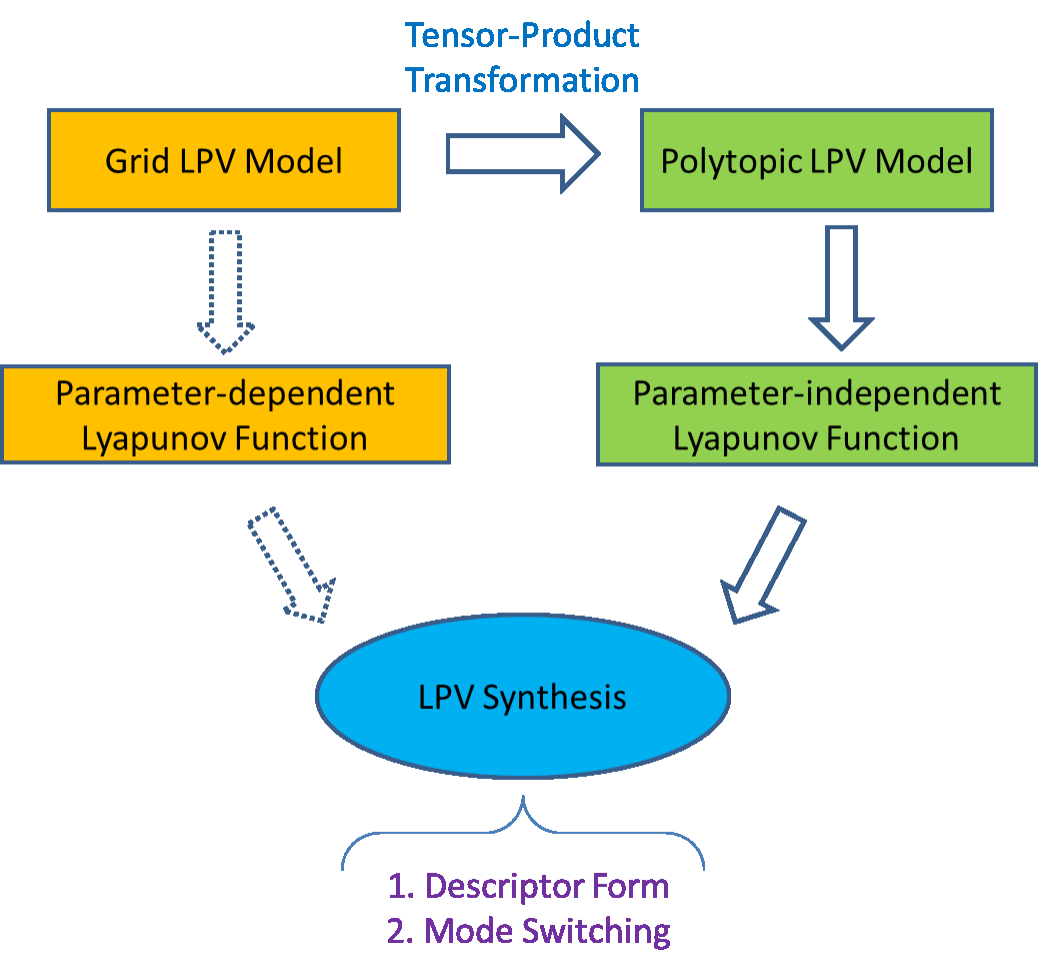}
\caption{LPV Model Conversion and Control Synthesis.}
\label{fig:PaperOrg}
\end{figure}
\section{LPV-based A/C Modelling}

\subsection{Jacobian Linearization}
The methods of the conversion from a nonlinear model to a LPV model can be classified into two categories: \emph{analytical} methods based on the availability of reliable nonlinear equations for the dynamics of the plant, from which suitable control-oriented representations can be derived; \emph{experimental} methods based entirely on identification, i.e., aiming at deriving LPV models for the plant directly from input/output data \cite{50_casella2008}. Since a control-oriented A/C plant model is available, the analytical method is pursued.  Three LPV modeling techniques belonging to analytical methods are widely adopted in practical applications, namely Jacobian linearization, state transformation and function substitution \cite{50_marcos2004}. In Jacobian linearization approach, a family of linear, time-invariant (LTI) plants at different points of interest throughout the operational envelope in order to obtain an LPV model. It is based on first-order linear approximations with respect to a set of equilibrium points.  Other modeling approaches involves the use of linear fractional transformations, velocity based approaches, or different types of linearizations. As the first application of LPV control technique on vapor compression cycle, the basic approach for LPV model derivation, namely Jacobian linearization, is selected to yield a grid LPV model.

During Jacobian linearization of the nonlinear A/C plant model, the envelop compassing trim points is determined by the final control objective. Generally, a controller is designed to track prescribed trajectories of two output variables, namely the pressure difference $\Delta p$ between the condenser and the evaporator, and the superheat temperature $SH$ at the evaporator, by the two actuators, namely the compressor speed $N_c$ and valve opening percentage $\alpha$. The underlying reason is that the evaporator pressure $p_e$ is an indicator of the cooling capacity of the system, and superheat temperature $SH$ impose a safety threshold of guaranteeing no liquid refrigerant entering the compressor. As the compressor rotation speed $N_c$ increases, the evaporator pressure $p_e$ drops to meet stronger cooling demand; as the valve $\alpha$ opens wider, the superheat temperature $SH$ becomes less as a result of superheated phase region length shrink. In other words, an one-to-one mapping exists between the envelope formed by the controlled inputs and the envelop formed by the regulated outputs.

\begin{figure*}[!htb]
  \centering
  \includegraphics[width=0.7\textwidth]{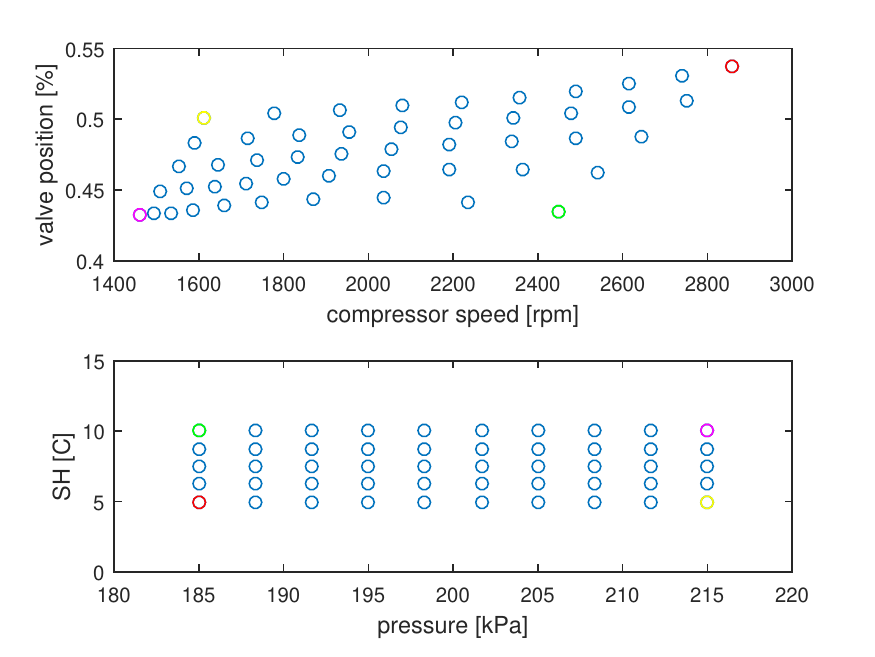}\\
  \caption{Linearization Trajectory}\label{F:Trajectory_Linearization}
\end{figure*}

For convenience, the envelope of the controlled inputs are sampled first in order to find appropriate envelop of the regulated outputs. In Figure \ref{F:Trajectory_Linearization}, eighteen trim points are selected to grid the trajectory inside a wider operational envelop. The corresponding operational envelop represented by the evaporator pressure and superheat temperature is in a rectangular. Along the six points selected for each of the three trajectories with fixed superheat temperature. The MATLAB function, \emph{trim}, is used to trim the nonlinear A/C plant model, and the MATLAB function, \emph{linmod}, is used to emulate Jacobian linearization technique.

The first step is the discretization of system matrices. Recall that the entire working envelope of the A/C system is that the pressure and superheat temperature varying from $180$ kPa to $220$ kPa and $0$ C to $15$ C, respectively. hence, the transformation space is defined as $\Omega = [180, 220] \times [0, 15]$. The gridding process is applied to the grid LPV model with a new density of the sampling grid as $[30,5]$, and the system matrices at the new sampling grid is interpolated with the old Jacobian linearization grid.

\subsection{Tensor Product Decomposition}
Next step is to determine the LTI vertex systems, $S_r$, and the weighting function, $w_r(p(t))$ using TP toolbox developed in \cite{50_nagy2007}. It is a numerical method that is capable of uniformly transforming LPV dynamic models into polytopic forms.  The TP model transformation generates two kinds of polytopic models. Firstly, it numerically reconstructs the HOSVD (Higher Order Singular Value) based canonical form of LPV models. This is a new and unique polytopic representation. This form extracts the unique structure and various important properties of a given LPV model in the same sense as the HOSVD does for matrices and tensors. Secondly, the TP model transformation generates various convex polytopic forms, upon which LMI (Linear Matrix Inequality) based multi-objective control design techniques can immediately be executed in order to satisfy the given control performance requirements. Here, the weighting type of \emph{cno convex} hull during the transformation in order to have a tight hull representation. The size of the tensor $S$ obtained is $20 \times 115 \times 4 \times 5$, indicating that the grid LPV model of the A/C plant can exactly be given as convex combination of $20 \times 115 = 2300$ LTI vertex system.  The singular values in the dimension of evaporator pressure and superheat temperature are given as:

$$
 [{HOSV_{SH}}]_{20 \times 1} =
       \left(
         \begin{array}{c}
           571.656  \\
           46.4382 \\
           4.38904 \\
           0.168789 \\
           ...  \\
           3.65281 \times 10^{-6}
         \end{array}
       \right),$$
$$
   [{HOSV_{Pe}}]_{115 \times 1} =
       \left(
         \begin{array}{c}
           571.328 \\
           47.7948 \\
           11.0925\\
           5.41856 \\
           ... \\
           1.0052 \times 10^{-8}
         \end{array}
       \right).
$$

However, in practice, the small numbers of the controllers are preferred for implementation in real application. In the third step, we kept the only three and three largest singular values in the dimension of evaporator pressure and superheat temperature, respectively.

\begin{figure*}
    \centering
      \subfigure[$w_{1,j}(SH(t))$]
      {\includegraphics[width=0.46\textwidth,keepaspectratio=true]{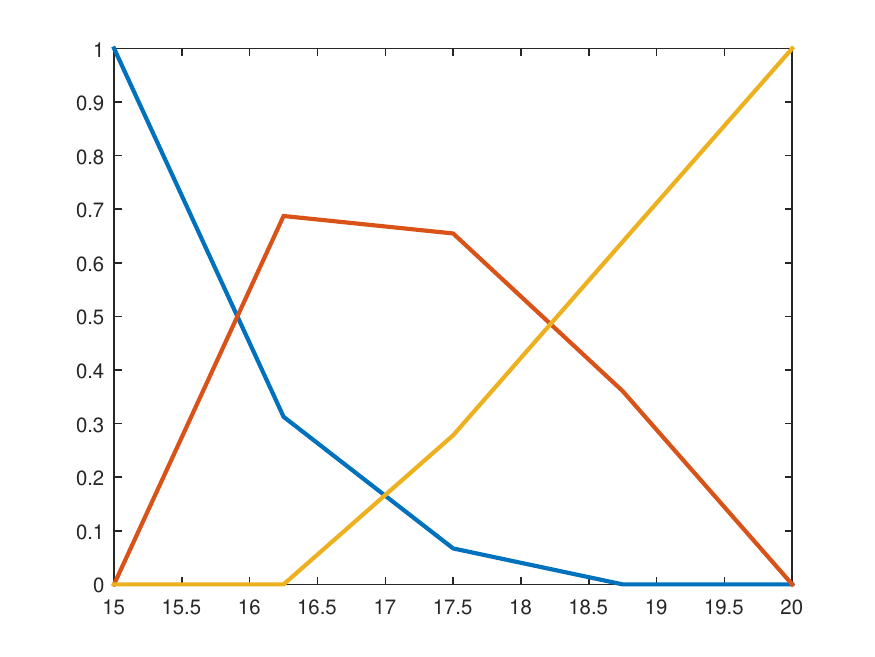}
      \label{F:SH_WeightingFunction}}
      \subfigure[$ w_{2,i}(p_e(t))$]
      {\includegraphics[width=0.46\textwidth,keepaspectratio=true]{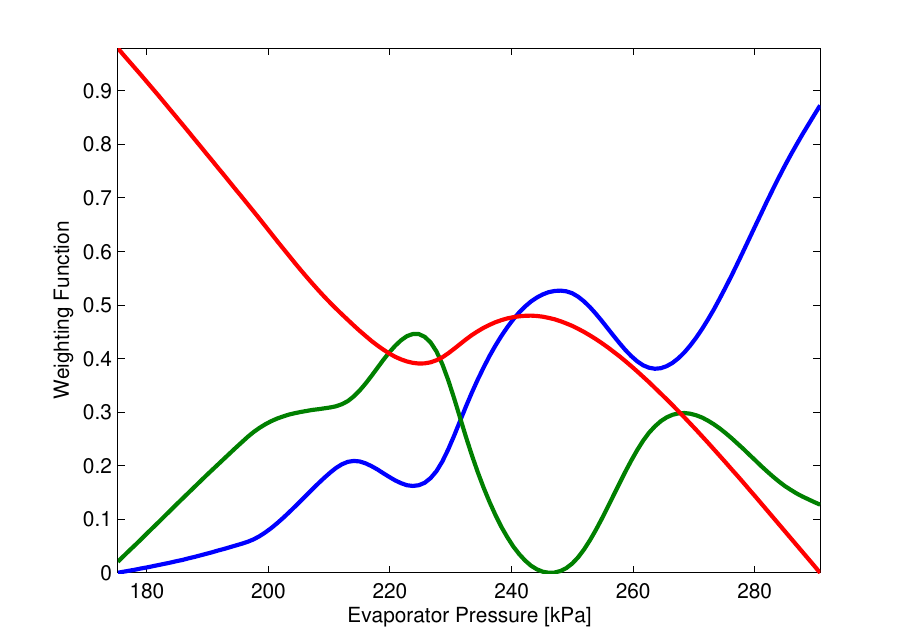}
      \label{F:Pe_WeightingFunction}}
      \caption{Weighting Functions in $w_{n,j}(p_n(t))$.}
      \label{F:WeightingFunction1}
\end{figure*}


\begin{figure*}[!htb]
      \centering
      {\includegraphics[width=0.85\textwidth,keepaspectratio=true]{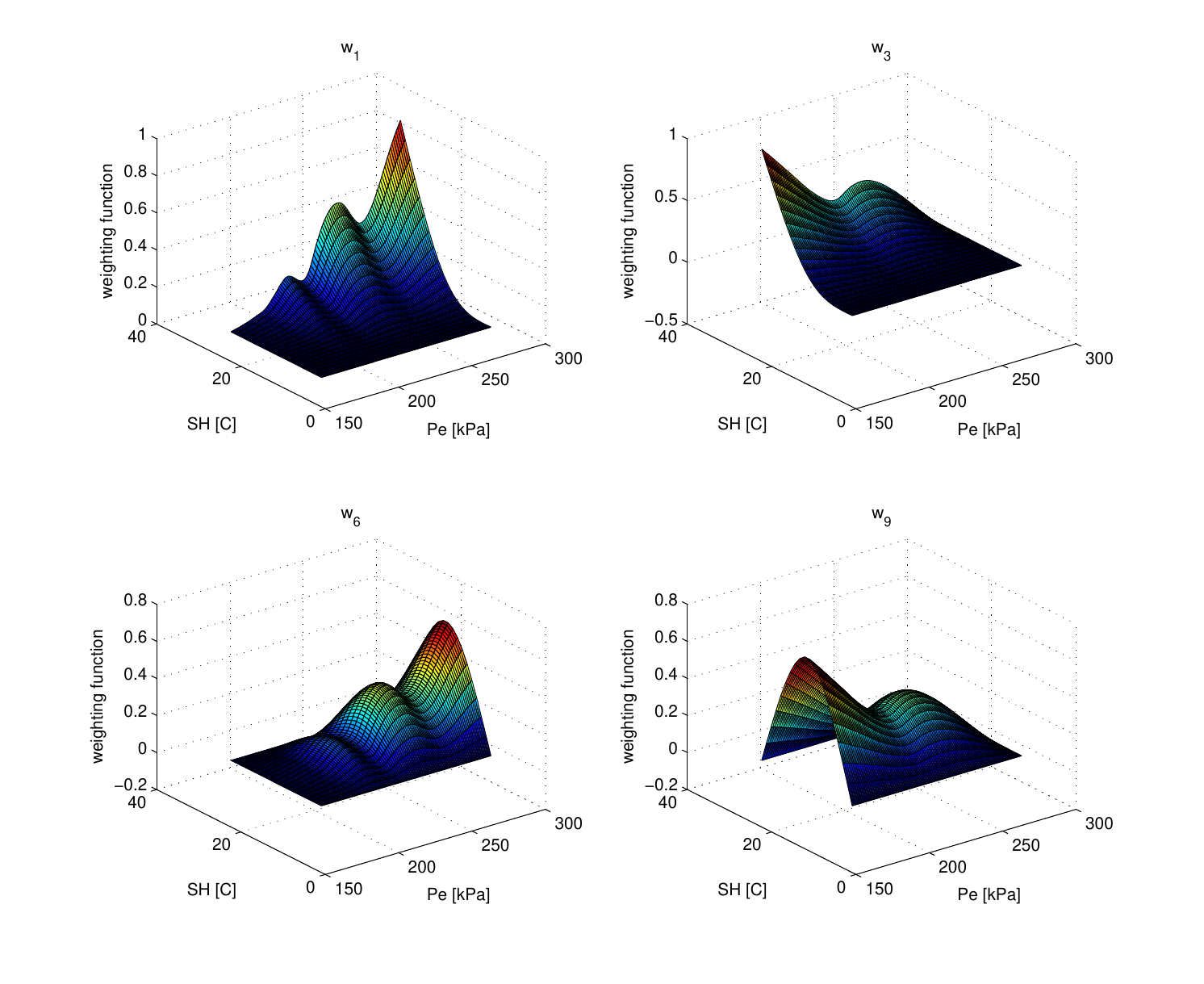}
      \caption{Weighting Functions in $w_{r}(p(t))$.}
      \label{F:TP_WeightingFunction}}
\end{figure*}

\subsection{LPV Formulation}
The LTI vertex systems was reduced to $3 \times 3=9$. Hence, the TP polytopic model can be written as
\begin{equation}
\begin{aligned}
   \dot{x}(t) & = \sum_{i=1}^{3}\sum_{j=1}^{3} w_{1,i}(SH(t)) w_{2,j}(p_e(t))(A_{i,j}x(t) + B_{i,j}u(t))\\
   y(t)       & = \sum_{i=1}^{3}\sum_{j=1}^{3} w_{1,i}(SH(t)) w_{2,j}(p_e(t))(C_{i,j}x(t) + D_{i,j}u(t))
\end{aligned}
\end{equation}
where the mapping between weighting functions and vertex systems are:
\begin{equation}
\left(
  \begin{array}{c|ccc}
      & w_{2,1} & w_{2,2} & w_{2,3} \\ \hline
    w_{1,1} & w_1 & w_2 & w_3 \\
    w_{1,2} & w_4 & w_5 & w_6 \\
    w_{1,3} & w_7 & w_8 & w_9 \\
  \end{array}
\right)
\Leftrightarrow
\left(
  \begin{array}{c|ccc}
      &  &  & \\ \hline
     & S_1 & S_2 & S_3 \\
    & S_4 & S_5 & S_6 \\
     & S_7 & S_8 & S_9 \\
  \end{array}
\right)
\end{equation}

The weighting function $w_{n,j}(p_n(t))$ are presented in Figure \ref{F:SH_WeightingFunction} and \ref{F:Pe_WeightingFunction}. Moreover, Figure \ref{F:TP_WeightingFunction} shows $w_{1}(p(t))$ , $w_{3}(p(t))$, $w_{6}(p(t))$, $w_{9}(p(t))$  as an example for determining $w_r(p(t))$. Since only two scheduling variables are available here, the weighting function for each vertex $w_r(p(t))$ is a product of the independent weighting function for each scheduling variables $w_{n,j}(p_n(t))$, for instance, $w_{1}(SH,p_e) = w_{1,1}(SH) \times w_{1,1}(p_e)$.

\begin{figure}[!htb]
  \centering
  \includegraphics[width=0.8\columnwidth]{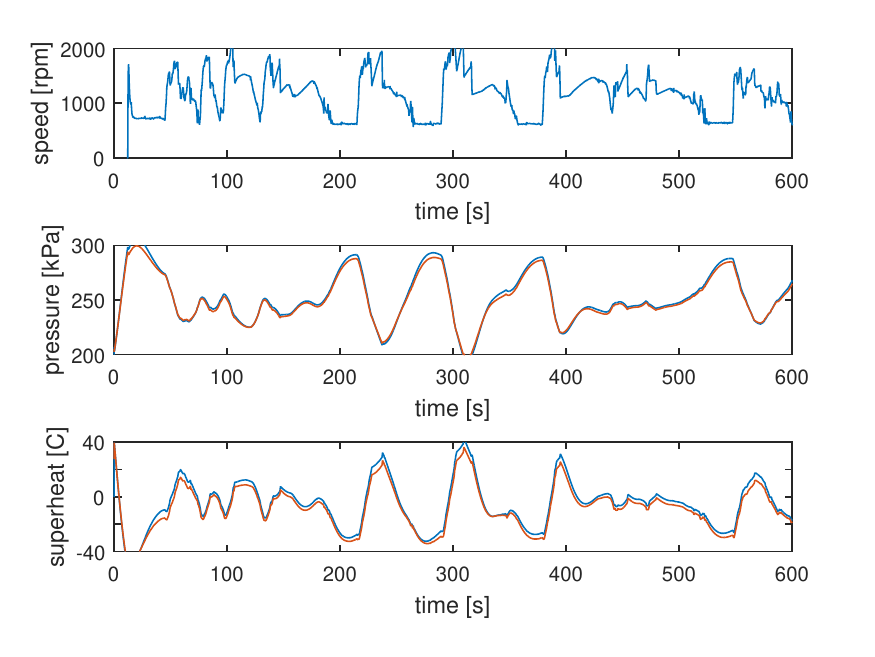}\\
  \caption{Compare Tensor Product Polytopic Model to Linear Parameter Varying Model}\label{F:CMP_Linear_TP}
\end{figure}

The accuracy of the decomposed TP polytopic model is guaranteed by comparison with the original grid LPV model. One approach is to test over plenty of points of randomly selected parameter values \cite{50_petres2006,50_petres2007}. Alternatively, the responses in time domain of both models are compared with different truncation of singular values during TP transformation. {\color{red}As shown in Figure \ref{F:CMP_Linear_TP}, the responses of the TP polytopic model with only the first two singular values kept are quite different from the grid LPV model. The response of the TP polytopic model with the first four singular values kept are almost the same as the one with the first three singular values kept.} Hence, accuracy is claimed to be maintained during the transformation from grid LPV model into TP polytopic LPV model.

\section{LPV Control Synthesis}

\subsection{Control Design}
\begin{figure*}[!htb]
  \centering
  \includegraphics[width=0.85\textwidth]{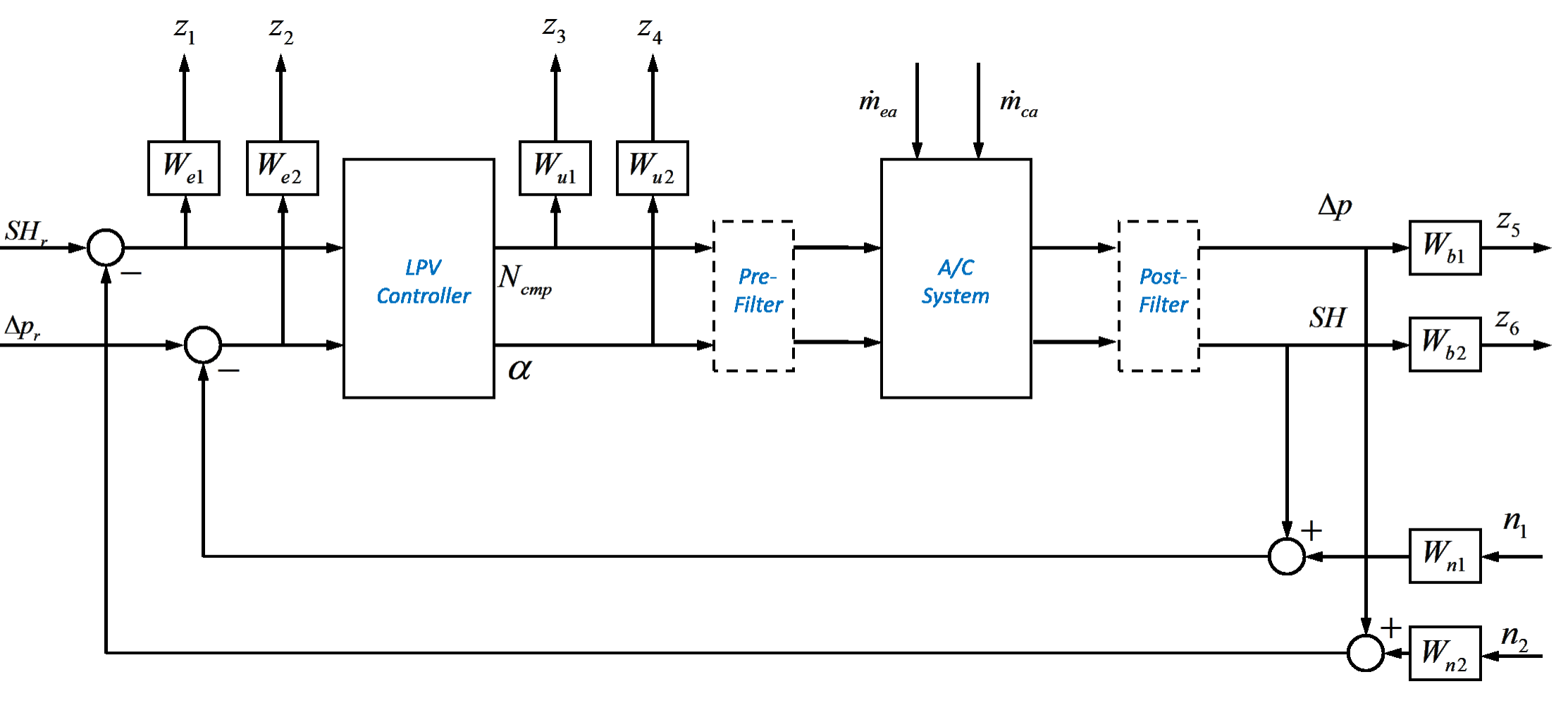}\\
  \caption{Typical A/C Configuration}\label{F: Control_Scheme}
\end{figure*}

In the A/C system, An LPV controller is designed to track prescribed trajectories of two output variables, namely the pressure difference $\Delta p$ between the condenser and the evaporator, and the superheat temperature $SH$ at the evaporator. The reference values for the tracked variables are indicated with $\Delta p_r$ and $SH_r$, respectively. At the same time, the controller should reject disturbances caused by air mass flow rate at the condenser, $\dot m_{ca}$, and the evaporator, $\dot m_{ea}$. Because there are no sensors mounted on vehicles for the two variables, the two disturbances are not measurable. They typically vary considerably in operation, due to the presence of fans and blowers, and the relative wind speed when the vehicle is in motion. In addition, noises $n_1$ and $n_2$ are present in the measured signals of measured superheat temperature $SH$ and pressure difference $\Delta p$. The interconnection between relevant blocks are shown in Figure \ref{F: Control_Scheme}.

The TP polytopic A/C plant model must be augmented for control algorithm development, especially the controlled output vector $z$ and the disturbance vector $\omega$.  The six elements in the output vector $z$ to be minimized are selected as
\begin{equation}
    \begin{bmatrix}
    e_{\Delta p} & e_{SH} & N_{cmp} & \alpha & \Delta p & SH
    \end{bmatrix}
    ^T,
\end{equation}
and the reasons of choosing each variable are given below:

\begin{enumerate}
  \item the errors $e_{\Delta p} = \Delta p_r - \Delta p$ and $e_{SH} = SH_r - SH$ on the output evaporator pressure and superheat should be minimized to achieve good tracking performance;

  \item the compressor rotation speed $N_{cmp}$ and the valve opening $\alpha$ should be varied as little as possible, resulting in less deviation from the nominal operating point of the A/C system and minimum control effort;

  \item the fluctuations in the pressure difference $\Delta p$ and superheat temperature $SH$ during transients should be limited to ensure stability of the A/C system model.
\end{enumerate}

Since the LPV controller design focuses on the frequency domain responses, weighting functions are added for improving the closed-loop performance.  After adding the weighting functions, the closed-loop output vector $z$ is defined as:
\begin{equation}
    \left(
      \begin{array}{c}
        z_1 \\
        z_2 \\
      \end{array}
    \right)
    =
    \left(
      \begin{array}{cc}
        \frac{K_{e1}}{s+\epsilon_{e1}} & 0 \\
        0 & \frac{K_{e2}}{s+\epsilon_{e2}} \\
      \end{array}
    \right)
    \left(
      \begin{array}{c}
        \Delta p_r - \Delta p \\
        SH_{r} - SH \\
      \end{array}
    \right)
\end{equation}

\begin{equation}
    \left(
      \begin{array}{c}
        z_5 \\
        z_6 \\
      \end{array}
    \right)
    =
    \left(
      \begin{array}{cc}
        \frac{K_{y1} s}{\epsilon_{y1} s + \omega_{y1}} & 0 \\
        0 & \frac{K_{y2} s}{\epsilon_{y2} s + \omega_{y2}}  \\
      \end{array}
    \right)
    \left(
      \begin{array}{c}
        \Delta p \\
        SH \\
      \end{array}
    \right)
\end{equation}
where the parameters of the weighting functions are selected as $ K_{e1} = 200, K_{e2} = 100,  \epsilon_{e1} = 400, \epsilon_{e2} = 800, K_{y1} = K_{y2} = 1, \epsilon_{y1}= \epsilon_{y2} = 0.1, \omega_{y1} = \omega_{y2} = 1$ \cite{05_zhou1998}.

The reference pressure difference $\Delta p_r$ and superheat temperature $SH_{r}$ are time-varying and regarded as additional disturbances besides the unknown disturbances $\dot m_{ca}$ and $\dot m_{ea}$, as well as the noises. Therefore, the disturbance vector is defined as:
\begin{equation}
    \omega = [\Delta \dot{m}_{ea}, \Delta \dot{m}_{ca}, \Delta p_r, SH_r ,n_1, n_2]
\end{equation}

Note that the augmented LPV plant model can also be obtained by describing the interconnection in Figure \ref{F: Control_Scheme} using the MATLAB function as \emph{connect}. The \emph{hinfgs} function in MATLAB Robust Control Toolbox \cite{50_gahinet1994} has given the set of LTI controllers for each vertex of the TP type convex polytopic form of the LPV A/C model. In order to adapt the augmented model to the requirement of the command, a pre-filter and a post-filter have been added to remove the parameter dependency of the matrix $B_1$, because it is a requirement of the controller synthesis.

The LPV controller is constructed by the combination of the vertex system matrices and weighting functions in the same fashion as the TP polytopic model. Hence,

\begin{equation}
\begin{aligned}
   \dot{x}_K(t) & \!=\! \sum_{i=1}^{3}\sum_{j=1}^{3} w_{1,i}(SH(t)) w_{2,j}(p_e(t))\\
   &\left(A_{i,j}x_K(t) \!+ \!B_{i,j}
   \left(
     \begin{array}{c}
       Pe_r - Pe \\
       SH_r - SH \\
     \end{array}
   \right)\right)
   \\
   \left(
     \begin{array}{c}
     N_c \\
     \alpha \\
     \end{array}
   \right)
          & = \sum_{i=1}^{3}\sum_{j=1}^{3} w_{1,i}(SH(t))w_{2,j}(p_e(t))\\
          &\left((C_{i,j}x_K(t) + D_{i,j}
   \left(
     \begin{array}{c}
       Pe_r - Pe \\
       SH_r - SH \\
     \end{array}
   \right)\right).
\end{aligned}
\end{equation}

\subsection{Performance Comparison}

The designed $H_{\infty}$ gain scheduling controller is validated with the nonlinear model of the A/C plant in MATLAB/Simulink. Note that the interpolation of the controller system matrices are realized automatically with the command \emph{tprod} in the TP transformation toolbox \cite{50_petres2006,50_petres2007}.

\begin{figure}[!htb]
  \centering
  \includegraphics[width=0.50\textwidth]{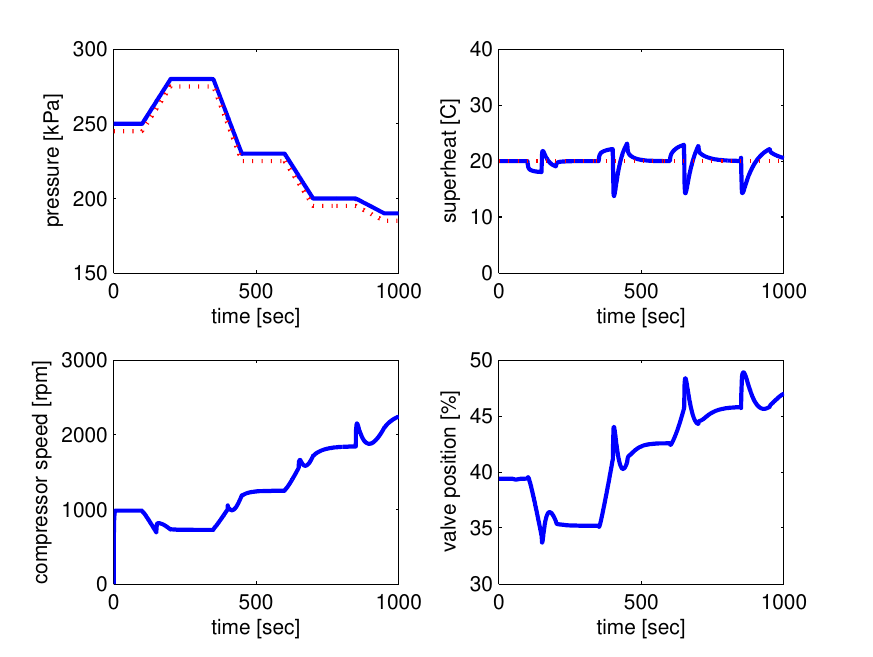}\\
  \caption{System Inputs and Outputs during Global Tracking without Disturbance}\label{F:sim_LPV}
  \centering
  \includegraphics[width=0.50\textwidth]{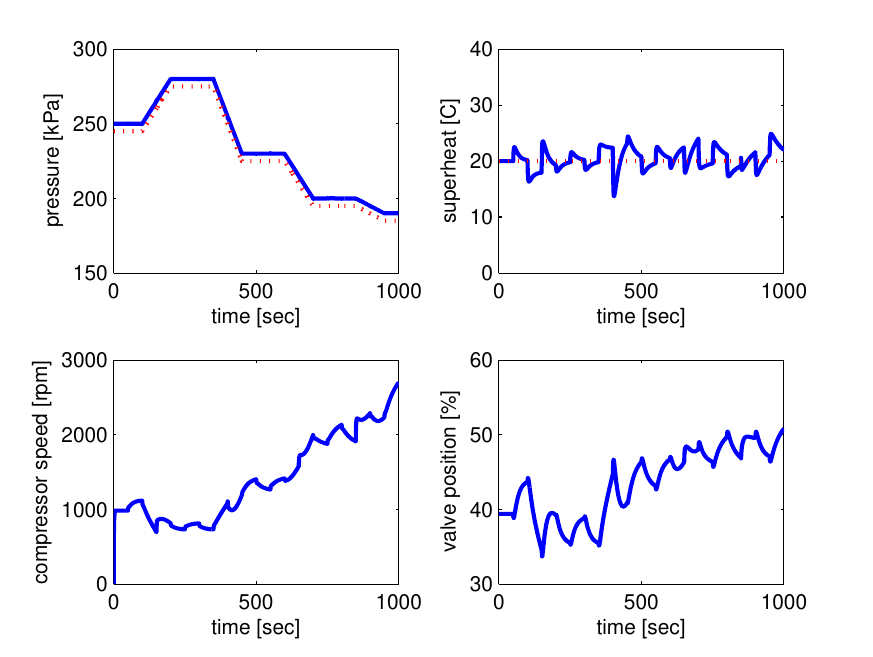}\\
  \caption{System Inputs and Outputs during Global Tracking with Disturbance added}\label{F:sim_LPV_dist}
\end{figure}

Figure \ref{F:sim_LPV} shows a global output tracking over the entire working region. The reference evaporator pressure, starting from medium cooling load, switches to low cooling load first and back to high cooling load finally. Due to the fast responses of the controller over evaporator pressure, the actual evaporator pressure almost overlaps the reference values. On the other hand, the actual superheat temperature shows noticeable transitions as the cooling load changes, and its maximum deviation is always within the safety threshold. In order to demonstrate the robustness of the designed controller over external disturbance. The variation of the air mass flow rate at the evaporator exterior surface is modeled as a pulse with period $100$ second and width $50$ second. Figure \ref{F:sim_LPV_dist} shows the tracking performance over the same reference signals after disturbances are added into the nonlinear A/C model. As expected, the controller designed is proved to maintain its capability of tracking the evaporator pressure very well. Due to the existence of external disturbance, the variation of the actual superheat temperature becomes more frequently than before, but still within the safety threshold. Therefore, it is demonstrated that tracking and robustness are both achieved over the entire working region.

\section{Open Issues}
Although the problem of output tracking of automotive AC systems has been adapted in the general framework of linear parameter varying modeling, identification and control, the designed controllers still suffer some constraints.

\subsection{Descriptor LPV Control}
Eliminate intermediate mass flow rates, we have

\begin{equation}
\label{E:evap_3}
\begin{aligned}
  &\left(\frac{\rho_{e,TP}-\rho_g}{\rho_{e,TP}}\right)\frac{d\zeta_1}{dt} + \frac{1}{\rho_{e,TP}}\frac{\partial \rho_{e,TP}}{\partial p_e}\frac{dp_e}{dt}\cdot\zeta_1 + \frac{1}{\rho_{e,TP}}\frac{\partial \rho_{e,TP}}{\partial \bar\gamma_e}\frac{d\bar\gamma_e}{dt}\cdot\zeta_1\\
   &= \frac{\dot m_v}{\rho_{e,TP}V_e} - \frac{\dot m_{12}}{\rho_{e,TP}V_e}\frac{\rho_g \left(h_{e,TP}-h_g\right)}{\rho_{e,TP}}\frac{d\zeta_1}{dt} + \left(\frac{\partial h_{e,TP}}{\partial p_e}-\frac{1}{\rho_{e,TP}}\right)\frac{dp_e}{dt}\cdot\zeta_1\\
   & + \frac{\partial h_{e,TP}}{\partial \bar\gamma_e}\frac{d\bar\gamma_e}{dt}\cdot\zeta_1  \\
  & = \frac{\dot m_v}{\rho_{e,TP}V_e} \left(h_4-h_{e,TP}\right) - \frac{\dot m_{12}}{\rho_{e,TP}V_e} \left(h_g-h_{e,TP}\right) +\frac{\dot{Q}_{TP}}{\rho_{e,TP}V_e}
\end{aligned}
\end{equation}

\begin{equation}
\label{E:evap_4}
\begin{aligned}
  &-\left(\frac{\rho_{e,SH}-\rho_g}{\rho_{e,SH}}\right)\frac{d\zeta_1}{dt} + \frac{1}{\rho_{e,SH}}\frac{\partial \rho_{e,SH}}{\partial p_e}\frac{dp_e}{dt}\cdot\left(1-\zeta_1\right)\\
   &+ \frac{1}{\rho_{e,SH}}\frac{\partial \rho_{e,SH}}{\partial h_{e,SH}}\frac{dh_{e,SH}}{dt}\cdot\left(1-\zeta_1\right) = \frac{\dot m_{12}}{\rho_{e,SH}V_e} - \frac{\dot m_{c}}{\rho_{e,SH}V_e}\\
  &-\frac{\rho_g \left(h_{g}-h_{e,SH}\right)}{\rho_{e,SH}}\frac{d\zeta_1}{dt} + \frac{1}{\rho_{e,TP}}\frac{dp_e}{dt}\cdot\left(1-\zeta_1\right) - \frac{dh_{e,SH}}{dt}\cdot\left(1-\zeta_1\right)  \\
  & = \frac{\dot m_{12}}{\rho_{e,SH}V_e} \left(h_g-h_{e,SH}\right) - \frac{\dot m_{c}}{\rho_{e,SH}V_e} \left(h_1-h_{e,SH}\right) +\frac{\dot{Q}_{SH}}{\rho_{e,SH}V_e}
\end{aligned}
\end{equation}

In the considered waste heat recovery system, the material thermodynamic properties are mathematically represented \cite{100_feru2014a,100_feru2014b}. The system can be written into rational form. The temperature-enthalpy characteristic of the working fluid for different values of the pressure is presented. Constants are used for the temperature-enthalpy characteristic on the liquid region. In the two-phase region the working fluid temperature is constant and equal to the saturation temperature. The vapor region is approximated using pressure dependent coefficients.  Also, the working fluid density as a function of enthalpy and pressure is illustrated.
\begin{equation}\label{E:affine}
    \rho_f = \left\{
      \begin{array}{ll}
        a_{\rho_l} h_f^2 + b_{\rho_l} h_f + c_{\rho_l}  & \hbox{if $h_f \leq h_l$;}\\
        TBD &  \hbox{if $h_l \leq h_f \leq h_v$;}\\
        a_{\rho_v} h_f^2 + b_{\rho_v} h_f + c_{\rho_v}  & \hbox{if $h_f \geq h_v$;}
      \end{array}
    \right.
\end{equation}

Note that change-of-variables method to descriptor systems has been shown \cite{100_rehm2000,100_rehm2002} with $E = diag\{I, 0\}$; below we provide formulas with general E and LMIs for a root clustering condition on descriptor systems.  Let us first consider LPV systems in the state space form,
where all coefficient matrices are rational parameter-dependent of scheduling variable $\theta$. Then via simple manipulations one can always derive a descriptor system that has the same input-output mapping and has coefficient matrices that are affine functions of $\theta$. Below is the representation of a LPV descriptor system
\begin{equation}\label{E:affine}
    \left\{
      \begin{array}{ll}
        E\dot{x}(t) = A x(t) + B_{1} \omega(t) + B_{2}u(t) \\
        z(t) = C_{1} x(t) + D_{11} \omega(t) + D_{1}u(t) \\
        y(t) = C_{2} x((t) + D_{21} \omega(t)
      \end{array}
    \right.
\end{equation}
Without loss of generality, we assume $E = diag\{I_r, 0\}$. 
Under the following assumptions: firstly, $B_{12} = 0$ and $C_{12} = 0$, secondly, either $B_{22} = 0$ or $C_{22} = 0$. The first assumption is satisfied if non affine functions do not appear in $B_{12}$ and $C_{1s}$, which are related to the external input and the controlled output, retrospectively. The second assumption means that the descriptor form does not have a hidden feedthrough term form the control input to the measured output. These assumptions are satisfied in the design of flight vehicle control presented later.

In \cite{100_masubuchi1997,100_masubuchi2003}, for the purpose of analysis and synthesis of gain-scheduling control systems, the descriptor form was used to represent LPV systems that have coefficient matrices of rational functions of the parameter if represented by the state space LPV model. It can be easily seen that through trivial augmentation of the descriptor form one can transform rational function valued coefficient matrices in the state space representation into a descriptor representation with an affine-function-valued coefficient. Based on the conventional LMI-based methods for descriptor systems, gain scheduling control system analysis and synthesis are formulated via parameter-dependent LMIs. 


In the work of \cite{100_polat2007}, the parametric LFT system is converted into descriptor form by addition of extra states. The descriptor matrix is constant and singular. This approach is distinct from our case, where the descriptor matrix is parametric and non-singular.

\subsection{Switched LPV Control}
The switching LPV control was introduced into the control of flight \cite{100_lu2004,100_lu2006}. In flight control, aircraft usually works in a wide angle of attack region. Designers often desire different performance goals in different angle of attack regions. For example, pilots need fast and accurate responses for maneuvering and attitude tracking in low angle of attack region, while in high angle of attack condition, the flight control emphasis lies in the maintainability of aircraft stability with acceptable flying qualities. The main idea in \cite{100_lu2004,100_lu2006} is to look for multiple Lyapunov functions to design different parameter-dependent controllers, and the hysteresis and average dwell time methods are proposed.

\begin{figure}
  \centering
  \subfigure[Hysteresis Switching Regions]
  {\includegraphics[width=0.75\columnwidth]{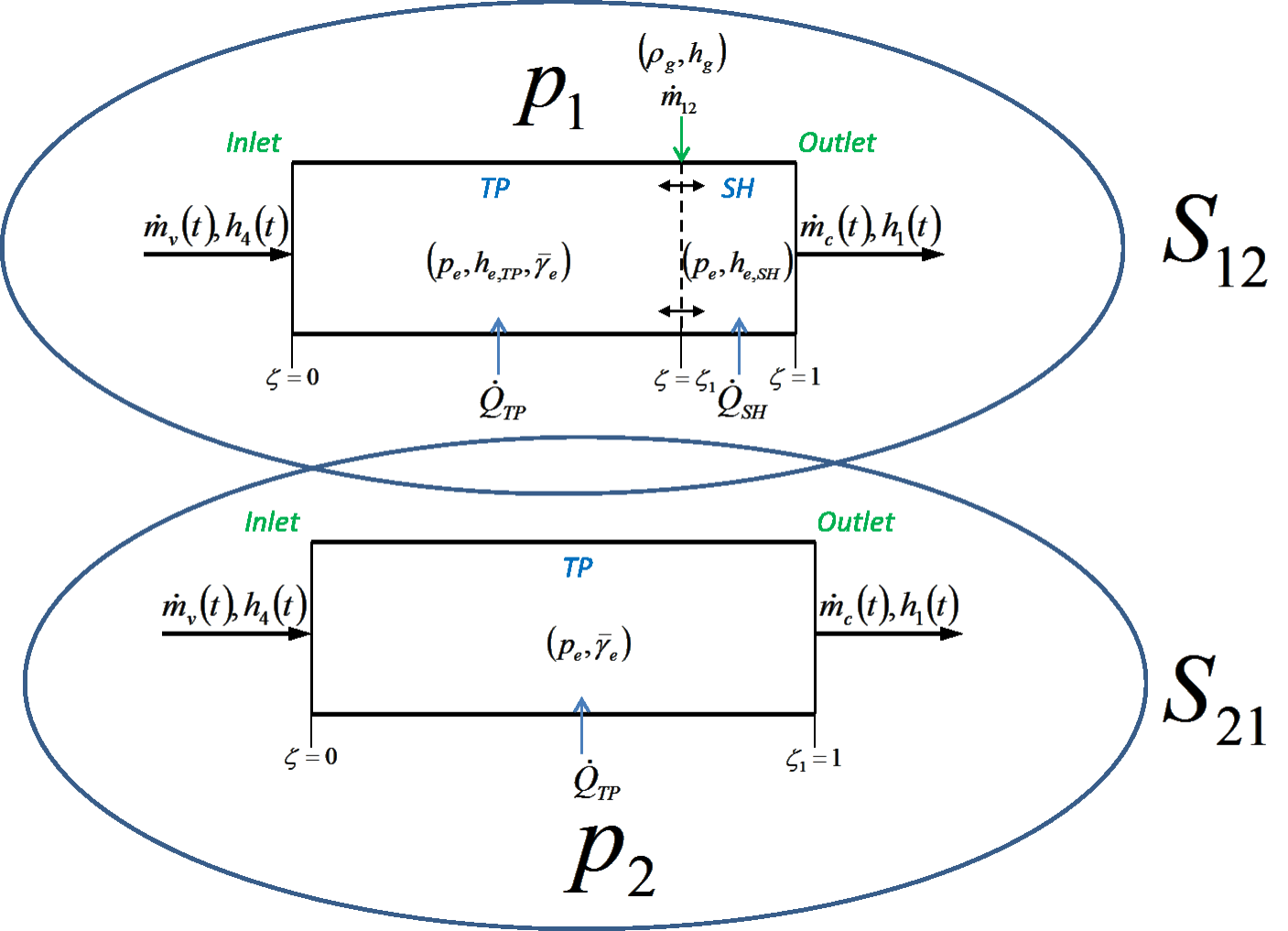}  \label{F:MassInbalance}}

  \subfigure[Switching Regions with Dwell Time]
  {\includegraphics[width=0.75\columnwidth]{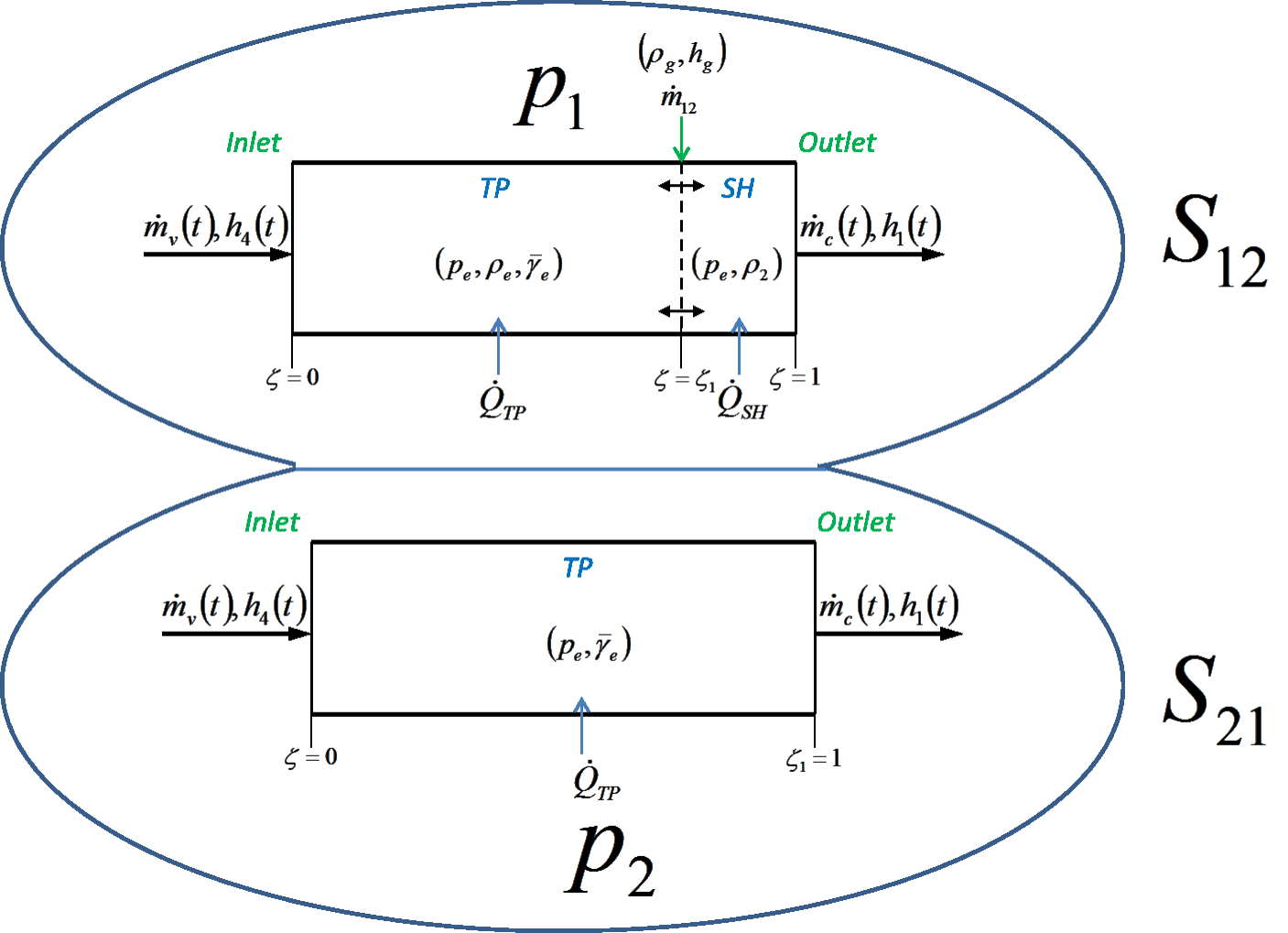}  \label{F:Massbalance}}
  \caption{}
\end{figure}

In order to capture the AC system dynamics during its start-up and shut-down, a multi-mode modeling approach is used to eliminate abrupt halt that might happen when some phase regions in the heat exchanger disappear and reappear. In an ideal vapor compression cycle, the inlet refrigerant to the evaporator is a saturated liquid-vapor mixture. The refrigerant leaving the evaporator ranges from a superheated vapor to a saturated liquid-vapor mixture through different operating conditions. Two different representations are needed to capture the evaporator dynamics during the stop-start cycle transients. These are evaporator mode 1 (two-phase and superheated two-zone model) and evaporator mode 2 (two-phase one-zone model) as illustrated in Figure \ref{F:MassInbalance} and \ref{F:Massbalance}. In \cite{02_McKinley2006, 02_Li2010}, the conditions to cause the switch from evaporator mode 1 to mode 2 depend on the state of the normalized length in the superheated zone, and are defined as
\begin{equation}\label{E:affine}
\begin{aligned}
    \zeta_{e2} < \zeta_{emin} \\
    \frac{d \zeta_{e2}}{dt} < 0
\end{aligned}
\end{equation}
The conditions are based on a minimum positive length of the superheated zone and on the sign of its derivative. The evaporator mode 2 switches to mode 2 when
\begin{equation}\label{E:affine}
\begin{aligned}
    \zeta_{e1} (\bar{\gamma}_e - \bar{\gamma}_{etot}) > \zeta_{emin}\\
    \frac{d \bar{\gamma}_e}{dt} > 0
\end{aligned}
\end{equation}

However, \cite{02_Cecchinato2011} the refrigerant pressure and superheated zone enthalpy are chosen as state variables. The choice is intrinsically mass conservative. This could be critical when simulating start-ups from low mean void fraction starting conditions; in fact for low evaporator outlet quality, the approximation error associated to the expansion of the two-phase zone mass and energy conservation equations is acceptable only within limited variations of the state variables given the strong sensitivity of void fraction to quality. A different approach introduces evaporator mean density as a state variable together with refrigerant pressure and superheated zone density, thus intrinsically ensuring refrigerant mass conservation during mode switching and low mean void fraction operation. This approach also ensures refrigerant energy conservation. A switch from mode 1 to mode 2 is triggered by the following conditions
\begin{equation}\label{E:affine}
\begin{aligned}
    \rho_2 \geq  \rho_g \\
    L_2 \leq 0
\end{aligned}
\end{equation}
This conditions can be described as the superheated zone density is higher than the vapor density and the superheated zone length is negative. The evaporator mode 2 switches to mode 2 when
\begin{equation}\label{E:affine}
\begin{aligned}
    \rho_2 < \rho_g \\
    L_2 > 0
\end{aligned}
\end{equation}

Suppose that there exist a family of positive definite matrix function ${X_i(\rho)}_{i \in Z_N}$, and each of them is smooth over the corresponding parameter subsect $P_i$. The multiple parameter-dependent Lyapunov functions can then be defined as
\begin{equation}
    V_{\sigma} (x_{cl},\rho) = x_{cl}^T X_{\sigma}(\rho)x_{cl}
\end{equation}
where the value of switching signal $\sigma$ represents the active operating region $P_{i}$ and thus determines the corresponding matrix function $X_i(\rho)$. Generally speaking, for a switched LPV system to be stable, the value of the discontinuous Lyapunov function $V_{\sigma}$ is not necessarily decreasing along the parameter trajectory. In fact, it is often enough to require that the value of $V_{\sigma}$ decreases in the active parameter region $P_i$ provided proper switching logic is adopted.  This will lead to relaxed stability conditions and provide enhanced control design flexibility.

When hysteresis switching logic is employed, it is assumed that any two adjacent parameter subsets are overlapped, as shown in Figure \ref{F:MassInbalance}. Thus, there are two switching surfaces between any two adjacent parameter subsets. We use $P_{ij}$ to denote the switching surface specifying the one directional move from subset $P_i$ to $P_j$. The switching event occurs when the parameter trajectory hits one of the switching surfaces $P_{ij}$ or $P_{ji}$. Theorem 1 in \cite{100_lu2004,100_lu2006} gives the sufficient conditions when the closed-loop LPV system is exponentially stabilized by switching LPV controllers over the entire parameters $P$, and its performance $\|e\|_2 \leq \gamma \|d\|_2 $ is achieved.

If the overlapped region between any two adjacent parameter subsets shrinks, it eventually becomes a single switching surface, as shown in Figure \ref{F:Massbalance}. Different from hysteresis switching, here $P_{ij}$ and $P_{ji}$ represent the same switching surface between subsets $P_i$ and $P_j$ regardless which direction the parameter trajectory moving from. This usually requires the continuity of Lyapunov function is non-increasing condition is to be satisfied on the switching surface. To relax continuity requirement of Lyapunov functions across the switching surfaces, we will consider another switching logic with average dwell time. The idea relaxes the concept of dwell time, allowing the possibility of switching fast when necessary and then compensating for it by switching sufficiently slow later on. Theorem 3 gives the synthesis condition of switching LPV control with average dwell time.

\section{Conclusion}
A controller for the A/C system cooling capacity tracking is designed by solving LMIs conditions specified by a specific type of LPV control synthesis method. The solution is made possible by resorting to a recently developed technique that transforms a Jacobian-based LPV model to a TP-type convex polytopic model form. Hence, special attention is paid to the details of model type conversion. The nonlinear simulation shows that the desired performance and robustness objectives can be achieved across the working region.

The performance of the LPV controller might be further improved provided that an analytical LPV model of the A/C plant is available. Currently, the Jacobian linearization is adopted to obtain a grid LPV model. The Jacobian linearization is the most restrictive in terms of operational envelope because it requires the existence of trim points. The main disadvantage is the resulted LPV controller needs to switch as the trim point changes, which explains the peaks generated during global output tracking. A better approach is to write the nonlinear A/C plant into descriptor LPV form directly, whose output tracking design has not been fully exploited yet. Moreover, the reference trajectory of the cooling capacity in current investigation is intentionally designed to change slowing in order to leave the controller enough time to converge. An alternative method is to incorporate the variation rate of the scheduling parameter, namely the evaporator pressure, into the design process of the LPV controller.

\section*{Acknowledgment}
The work described in the paper is in part supported by the U.S. Department of Energy, through Chrysler, LLC as the prime contractor. The authors gratefully acknowledge Chrysler, LLC and Dr. Timothy C. Scott for providing the data to calibrate the model and for the useful discussions.

\section*{Appendix:Tensor Product Transformation}
The TP model transformation is an executable numerical method and has three key steps. The first step is the discretization of the given system matrix over a huge number of points $p(t) \in \Omega$, where the parameter space is $\Omega = [a_1, b_1] \times [a_2, b_2] \times \dots \times [a_N, b_N] $. The discretized points are defined by a dense hyper-rectangular grid. The discretized system matrices at each grid point are stored into a tensor.

The second step extracts the LTI vertex systems from the discretized systems using a high order singular value decomposition (HOSVD) to decompose a given N-dimensional tensor into a full orthonormal system in a special ordering of higher order singular values, expressing the rank properties of the tensor in the order of $L_2$ norm. Meanwhile, it defines the continuous weighting functions to the LTI vertex system. The mathematical details are available in  \cite{50_baranyi2004, 50_petres2007,50_petres2006}. The resulting  TP polytopic model in terms of tensor product is given as
\begin{equation}
   \left(
     \begin{array}{c}
       \dot{x}(t) \\
       y(t) \\
     \end{array}
   \right)
   =
   S \bigotimes_{n=1}^N W_n(p_n(t))
   \left(
     \begin{array}{c}
       x(t) \\
       u(t) \\
     \end{array}
   \right)
\end{equation}
where row vector $W_n(p_n(t)) \in R^{I_n}, n = 1, \dots, N$ contains one bounded variable, and continuous weighting functions $w_{n,j}, j=1,\dots,I_n$, and $I_n$ is the number of the weighting functions used in the $n$th dimension of the parameter vector $p(t)$. The weighting function $w_{n,j}(p_n(t))$ is the $j$th weighing function defined on the nth dimension of $\Omega = [a_1, b_1] \times [a_2, b_2] \times \dots \times [a_N, b_N] $, and $p_n(t)$ is the $n$th  element of vector $p(t)$. The tensor $S \in R^{I_1 \times I_2 \times \dots \times I_N \times O \times I}$ is constructed from the LTI vertex systems $S_r \in R^{O \times I}$, where $O = n+q$ and $I=n+m$, and

\begin{equation}
   S_r =
   \left(
     \begin{array}{cc}
       A_r & B_r \\
       C_r & D_r \\
     \end{array}
   \right)
   = S_{i_1 i_2 \dots i_N}
\end{equation}
where $r=ordering (i_1 i_2 \dots i_N)$, $r=1$ to R, and $R= \prod\limits_n I_n$. Note that $S_r \neq S_i$. Define the weighting functions according to the sequence of r
\begin{equation}
   w_r(p(t)) = \prod\limits_n w_{n,j}(p_n(t))
\end{equation}

The goal of the TP model transformation is to determine the LTI vertex systems, $S_r$, and the weighting functions, $w_{n,j}(p_n(t))$, such that the system matrix, $S(p(t))$, is given for any grid points over the entire parameter spaces, and can be expressed as the combination of the vertex system matrices, $S_r$, and the weighting function, $w_r(p(t))$, which are actually nonlinearly dependent on the time-varying parameters,
\begin{equation}
\begin{aligned}
   S \bigotimes_{n=1}^N W_n(p_n(t))
   = \sum_{i=1}^R w_r(p(t))S_r \\
   \|S(p(t)) - \sum_{i=1}^R w_r(p(t))S_r\| \leq \epsilon
\end{aligned}
\end{equation}
Here, $\epsilon$ symbolizes the approximation error. The error arises in the third step, where it discards all zero or small singular values and their corresponding singular vectors in all N-dimensions.

\bibliographystyle{plain}
\bibliography{ACbiblio}

\begin{thebibliography}{10}

\bibitem{50_apkarian1998}
Pierre Apkarian and Richard~J Adams.
\newblock Advanced gain-scheduling techniques for uncertain systems.
\newblock {\em Control Systems Technology, IEEE Transactions on}, 6(1):21--32,
  1998.

\bibitem{50_baranyi2004}
P{\'e}ter Baranyi.
\newblock Tp model transformation as a way to lmi-based controller design.
\newblock {\em Industrial Electronics, IEEE Transactions on}, 51(2):387--400,
  2004.

\bibitem{50_casella2008}
Francesco Casella and Marco Lovera.
\newblock Lpv/lft modelling and identification: overview, synergies and a case
  study.
\newblock In {\em Computer-Aided Control Systems, 2008. CACSD 2008. IEEE
  International Conference on}, pages 852--857. IEEE, 2008.

\bibitem{02_Cecchinato2011}
L.~Cecchinato and F.~Mancini.
\newblock An intrinsically mass conservative switched evaporator model adopting
  the moving-boundary method.
\newblock {\em International Journal of refrigeration}, 35(2):349--364, 2011.

\bibitem{50_chumalee2009}
Sunan Chumalee and James Whidborne.
\newblock Lpv autopilot design of a jindivik uav.
\newblock In {\em AIAA Guidance, Navigation, and Control Conference and
  Exhibit}, 2009.

\bibitem{7533442}
V.~S. Deshpande, P.~D. Shendge, and S.~B. Phadke.
\newblock Nonlinear control for dual objective active suspension systems.
\newblock {\em IEEE Transactions on Intelligent Transportation Systems}, 2016.
\newblock doi:10.1109/TITS.2016.2585343.

\bibitem{Du2005981}
Haiping Du, Kam~Yim Sze, and James Lam.
\newblock Semi-active ${H}_{\infty}$ control of vehicle suspension with
  magneto-rheological dampers.
\newblock {\em Journal of Sound and Vibration}, 283(3–5):981--996, 2005.

\bibitem{100_feru2014b}
Emanuel Feru, Bram de~Jager, Frank Willems, and Maarten Steinbuch.
\newblock Two-phase plate-fin heat exchanger modeling for waste heat recovery
  systems in diesel engines.
\newblock {\em Applied Energy}, 133:183--196, 2014.

\bibitem{100_feru2014a}
Emanuel Feru, Frank Willems, Bram de~Jager, and Maarten Steinbuch.
\newblock Modeling and control of a parallel waste heat recovery system for
  euro-vi heavy-duty diesel engines.
\newblock {\em Energies}, 7(10):6571--6592, 2014.

\bibitem{50_gahinet1994}
PM~Gahinet, Arkadii Nemirovskii, Alan~J Laub, and Mahmoud Chilali.
\newblock The lmi control toolbox.
\newblock In {\em IEEE Conference on Decision and Control}, volume~2, pages
  2038--2038. INSTITUTE OF ELECTRICAL ENGINEERS INC (IEE), 1994.

\bibitem{01_Asada1998}
X.~He, S.~Liu, H.~Asada, and H.~Itoh.
\newblock Multivariable control of vapor compression systems.
\newblock {\em HVAC\&R Research}, 4(3):205--230, 1998.

\bibitem{02_He1997}
X-D He, Sheng Liu, and Haruhiko~H Asada.
\newblock Modeling of vapor compression cycles for multivariable feedback
  control of hvac systems.
\newblock {\em Journal of dynamic systems, measurement, and control},
  119(2):183--191, 1997.

\bibitem{02_Jensen2003}
J.M. Jensen.
\newblock {\em Dynamic Modeling of Thermo-Fluid Systems}.
\newblock PhD thesis, Technical University of Denmark, Department of Energy
  Engineering, 2003.

\bibitem{04_leducq2006}
D.~Leducq, J.~Guilpart, and G.~Trystram.
\newblock Non-linear predictive control of a vapour compression cycle.
\newblock {\em International journal of refrigeration}, 29(5):761--772, 2006.

\bibitem{1219456}
Loo~Hay Lee, Kay~Chen Tan, Ke~Ou, and Yoong~Han Chew.
\newblock Vehicle capacity planning system: a case study on vehicle routing
  problem with time windows.
\newblock {\em IEEE Transactions on Systems, Man, and Cybernetics - Part A:
  Systems and Humans}, 33(2):169--178, March 2003.

\bibitem{02_Li2010}
B.~Li and A.G. Alleyne.
\newblock A dynamic model of a vapor compression cycle with shut-down and
  start-up operations.
\newblock {\em International Journal of refrigeration}, 33(3):538--552, 2010.

\bibitem{J8}
Z.~Li, D.~P. Filev, I.~Kolmanovsky, E.~Atkins, and J.~Lu.
\newblock A new clustering algorithm for processing gps-based road anomaly
  reports with a mahalanobis distance.
\newblock {\em IEEE Transactions on Intelligent Transportation Systems}, 2016.
\newblock doi:10.1109/TITS.2016.2614350.

\bibitem{C1}
Z.~Li, I.~Kolmanovsky, E.~Atkins, J.~Lu, and D.~Filev.
\newblock Road anomaly estimation: Model based pothole detection.
\newblock In {\em 2015 American Control Conference (ACC)}, pages 1315--1320,
  July 2015.

\bibitem{C2}
Z.~Li, I.~Kolmanovsky, E.~Atkins, J.~Lu, D.~Filev, and J.~Michelini.
\newblock Cloud aided safety-based route planning.
\newblock In {\em 2014 IEEE International Conference on Systems, Man, and
  Cybernetics}, pages 2495--2500, Oct 2014.

\bibitem{C3}
Z.~Li, I.~Kolmanovsky, E.~Atkins, J.~Lu, D.~Filev, and J.~Michelini.
\newblock Cloud aided semi-active suspension control.
\newblock In {\em Computational Intelligence in Vehicles and Transportation
  Systems (CIVTS), 2014 IEEE Symposium on}, pages 76--83, Dec 2014.

\bibitem{J3}
Z.~Li, I.~Kolmanovsky, E.~Atkins, J.~Lu, D.~P. Filev, and J.~Michelini.
\newblock Road risk modeling and cloud-aided safety-based route planning.
\newblock {\em IEEE Transactions on Cybernetics}, 46(11):2473--2483, Nov 2016.

\bibitem{J2}
Z.~Li, I.~V. Kolmanovsky, U.~V. Kalabic, E.~M. Atkins, J.~Lu, and D.~P. Filev.
\newblock Optimal state estimation for systems driven by jump-diffusion process
  with application to road anomaly detection.
\newblock {\em IEEE Transactions on Control Systems Technology}, 2016.
\newblock doi:10.1109/TCST.2016.2620062.

\bibitem{C9}
Z.~Li, H.~R. Ossareh, I.~V. Kolmanovsky, E.~M. Atkins, and J.~Lu.
\newblock Nonlinear control of semi-active suspension systems: A quasi-linear
  control approach.
\newblock In {\em 2016 American Control Conference (ACC)}, pages 2397--2402,
  July 2016.

\bibitem{C5}
Z.~Li, X.~Yin, I.~Kolmanovsky, J.~Lu, D.~Filev, and E.~Atkins.
\newblock Robust ${H}_{\infty}$ control for a class of networked uncertain
  systems with multiple channels subject to markovian switching.
\newblock In {\em 2015 54th IEEE Conference on Decision and Control (CDC)},
  pages 6856--6861, Dec 2015.

\bibitem{C4}
Zhaojian Li, Ilya Kolmanovsky, Ella Atkins, Jianbo Lu, and Dimitar Filev.
\newblock ${H}_{\infty}$ filtering for cloud-aided semi-active suspension with
  delayed road information.
\newblock {\em IFAC-PapersOnLine}, 48(12):275--280, 2015.
\newblock 12th \{IFAC\} Workshop on Time Delay Systems, Ann Arbor, Michigan.

\bibitem{7506101}
B.~Liu, M.~Saif, and H.~Fan.
\newblock Adaptive fault tolerant control of a half-car active suspension
  systems subject to random actuator failures.
\newblock {\em IEEE/ASME Transactions on Mechatronics}, 21(6):2847--2857, Dec
  2016.

\bibitem{100_lu2004}
Bei Lu and Fen Wu.
\newblock Switching lpv control designs using multiple parameter-dependent
  lyapunov functions.
\newblock {\em Automatica}, 40(11):1973--1980, 2004.

\bibitem{100_lu2006}
Bei Lu, Fen Wu, and SungWan Kim.
\newblock Switching lpv control of an f-16 aircraft via controller state reset.
\newblock {\em Control Systems Technology, IEEE Transactions on},
  14(2):267--277, 2006.

\bibitem{50_marcos2004}
Andr{\'e}s Marcos and Gary~J Balas.
\newblock Development of linear-parameter-varying models for aircraft.
\newblock {\em Journal of Guidance, Control, and Dynamics}, 27(2):218--228,
  2004.

\bibitem{100_masubuchi2003}
Izumi Masubuchi, Tomoaki Akiyama, and Masami Saeki.
\newblock Synthesis of output feedback gain-scheduling controllers based on
  descriptor lpv system representation.
\newblock In {\em Decision and Control, 2003. Proceedings. 42nd IEEE Conference
  on}, volume~6, pages 6115--6120. IEEE, 2003.

\bibitem{100_masubuchi1997}
Izumi Masubuchi, Yoshiyuki Kamitane, Atsumi Ohara, and Nobuhide Suda.
\newblock H∞ control for descriptor systems: a matrix inequalities approach.
\newblock {\em Automatica}, 33(4):669--673, 1997.

\bibitem{02_McKinley2006}
T.L. McKinley and A.G. Alleyne.
\newblock An advanced nonlinear switched heat exchanger model for vapor
  compression cycles using the moving-boundary method.
\newblock {\em International Journal of refrigeration}, 31(7):1253--1264, 2008.

\bibitem{50_nagy2007}
Szabolcs Nagy, Zolt{\'a}n Petres, and P{\'e}ter Baranyi.
\newblock Tp tool-a matlab toolbox for tp model transformation.
\newblock In {\em Proc. of 8th International Symposium of Hungarian Researchers
  on Computational Intelligence and Informatics}, pages 483--495, 2007.

\bibitem{50_petres2006}
Zolt{\'a}n Petres.
\newblock Polytopic decomposition of linear parameter-varying models by
  tensor-product model transformation.
\newblock {\em Budapest: Budapest University of Technology and Economics},
  2006.

\bibitem{50_petres2007}
Zolt{\'a}n Petres, P{\'e}ter Baranyi, P{\'e}ter Korondi, and Hideki Hashimoto.
\newblock Trajectory tracking by tp model transformation: Case study of a
  benchmark problem.
\newblock {\em Industrial Electronics, IEEE Transactions on}, 54(3):1654--1663,
  2007.

\bibitem{100_polat2007}
I~Polat, E~Eskinat, and IE~Kose.
\newblock Dynamic output feedback control of quasi-lpv mechanical systems.
\newblock {\em Control Theory \& Applications, IET}, 1(4):1114--1121, 2007.

\bibitem{50_rangajeeva2011}
S.~Rangajeeva and J.~Whidborne.
\newblock Linear parameter varying control of a quadrotor.
\newblock In {\em Industrial and Information Systems (ICIIS), 2011 6th IEEE
  International Conference on}, pages 483--488. IEEE, 2011.

\bibitem{04_rasmussen2010}
B.P. Rasmussen and A.G. Alleyne.
\newblock Gain scheduled control of an air conditioning system using the youla
  parameterization.
\newblock {\em IEEE Transactions on Control Systems Technology},
  18(5):1216--1225, 2010.

\bibitem{04_rasmussen2011}
Henrik Rasmussen and Lars Finn~Sloth Larsen.
\newblock Non-linear and adaptive control of a refrigeration system.
\newblock {\em IET Control Theory \& Applications}, 5(2):364--378, 2011.

\bibitem{100_rehm2000}
A~Rehm and F~Allg{\"o}wer.
\newblock Self-scheduled h∞ output feedback control of descriptor systems.
\newblock {\em Computers \& Chemical Engineering}, 24(2):279--284, 2000.

\bibitem{100_rehm2002}
Ansgar Rehm and Frank Allg{\"o}wer.
\newblock General quadratic performance analysis and synthesis of differential
  algebraic equation (dae) systems.
\newblock {\em Journal of Process Control}, 12(4):467--474, 2002.

\bibitem{04_shah2004}
R.~Shah, B.~P~Rasmussen, and A.G. Alleyne.
\newblock Application of a multivariable adaptive control strategy to
  automotive air conditioning systems.
\newblock {\em International Journal of Adaptive Control and Signal
  Processing}, 18(2):199--221, 2004.

\bibitem{6214702}
P.~B. Sujit, D.~E. Lucani, and J.~B. Sousa.
\newblock Bridging cooperative sensing and route planning of autonomous
  vehicles.
\newblock {\em IEEE Journal on Selected Areas in Communications},
  30(5):912--922, June 2012.

\bibitem{50_wu1996}
Fen Wu, Xin~Hua Yang, Andy Packard, and Greg Becker.
\newblock Induced l2-norm control for lpv systems with bounded parameter
  variation rates.
\newblock {\em International Journal of Robust and Nonlinear Control},
  6(9-10):983--998, 1996.

\bibitem{Yamashita19941717}
Masashi Yamashita, Kazuo Fujimori, Kisaburo Hayakawa, and Hidenori Kimura.
\newblock Application of ${H}_{\infty}$ control to active suspension systems.
\newblock {\em Automatica}, 30(11):1717--1729, 1994.

\bibitem{C7}
X.~Yin, Z.~Li, S.~L. Shah, L.~Zhang, and C.~Wang.
\newblock Fuel efficiency modeling and prediction for automotive vehicles: A
  data-driven approach.
\newblock In {\em Systems, Man, and Cybernetics (SMC), 2015 IEEE International
  Conference on}, pages 2527--2532, Oct 2015.

\bibitem{J1}
X.~Yin, Z.~Li, L.~Zhang, and M.~Han.
\newblock Distributed state estimation of sensor-network systems subject to
  markovian channel switching with application to a chemical process.
\newblock {\em IEEE Transactions on Systems, Man, and Cybernetics: Systems},
  2016.
\newblock doi:10.1109/TSMC.2016.2632155.

\bibitem{J6}
X.~Yin, Z.~Li, L.~Zhang, C.~Wang, W.~Shammakh, and B.~Ahmad.
\newblock Model reduction of a class of markov jump nonlinear systems with
  time-varying delays via projection approach.
\newblock {\em Neurocomputing}, 166:436--446, 2015.

\bibitem{J9}
X.~Yin, L.~Zhang, Z.~Ning, D.~Tian, A.~Alsaedi, and B.~Ahmad.
\newblock State estimation via {M}arkov switching-channel network and
  application to suspension systems.
\newblock {\em IET Control Theory \& Applications}, 2016.
\newblock doi:10.1049/iet-cta.2016.1108.

\bibitem{J7}
X.~Yin, L.~Zhang, Y.~Zhu, C.~Wang, and Z.~Li.
\newblock Robust control of networked systems with variable communication
  capabilities and application to a semi-active suspension system.
\newblock {\em IEEE/ASME Transactions on Mechatronics}, 21(4):2097--2107, Aug
  2016.

\bibitem{04_zhang20132}
Quansheng Zhang and Marcello Canova.
\newblock Lumped-parameter modeling of an automotive air conditioning system
  for energy optimization and management.
\newblock In {\em ASME 2013 Dynamic Systems and Control Conference}, pages
  V001T04A003--V001T04A003. American Society of Mechanical Engineers, 2013.

\bibitem{50_zhang2014JDSMC}
Quansheng Zhang and Marcello Canova.
\newblock Modeling and feedback control of a vehicle air conditioning system.
\newblock {\em Submitted to Journal of dynamic systems, measurement, and
  control}, 2014.

\bibitem{04_zhang2013}
Quansheng Zhang, Marcello Canova, and Giorgio Rizzoni.
\newblock Sliding mode control of an automotive air conditioning system.
\newblock In {\em American Control Conference}, pages 5748--5753. IEEE, 2013.

\bibitem{04_zhang2014}
Quansheng Zhang, Marcello Canova, and Giorgio Rizzoni.
\newblock Robust control of an automotive air conditioning system.
\newblock In {\em American Control Conference}, pages 5748--5753. IEEE, 2014.

\bibitem{05_zhou1998}
Kemin Zhou and John~Comstock Doyle.
\newblock {\em Essentials of robust control}, volume 104.
\newblock Prentice Hall Upper Saddle River, NJ, 1998.

\end{thebibliography}

\end{document}